# Widely Used and Fast De Novo Drug Design by a Protein Sequence-Based Reinforcement Learning Model

Yaqin Li[1], Lingli Li[1,*], Yongjin Xu[3,*], and Yi Yu[2,*]

[1]West China Hospital, Sichuan University, Chengdu 610041, China
[2]Department of Chemistry and Molecular Biology, University of Gothenburg, Kemivägen 10, 412 96 Gothenburg, Sweden.
[3]Department of Lymphoma, The Cancer Hospital of the University of Chinese Academy of Sciences (Zhejiang Cancer Hospital), Institute of Basic Medicine and Cancer (IBMC), Chinese Academy of Sciences, Hangzhou, PR China
[*] Author to whom any correspondence should be addressed.

**Abstract:** De novo molecular design has facilitated the exploration of large chemical space to accelerate drug discovery. Structure-based de novo method can overcome the data scarcity of active ligands by incorporating drug-target interaction into deep generative architectures. However, these strategies are bottlenecked by the small fraction of experimentally determined protein or complex structures. In addition, the cost of the molecular generation is computationally expensive due to 3D representations of both molecule and protein. Here, we demonstrate a widely used and fast protein sequence-based reinforcement learning (RL) model for drug discovery. In the generative model, one of the reward components, a binding affinity predictor, is based on 1D protein sequence and molecular SMILES. As a proof of concept, the RL model was utilized to design molecules for four targets. The generated compounds showed bioactivities by the validation of both QSAR and molecular docking with experimental 3D binding pockets. We also found that the performance of generated molecules depends on the selection of data source training for the binding predictor. Furthermore, drug design for a kinase without any experimental structure, CDK20, was studied. By only 1D protein sequence as input, the generated novel compounds showed favorable binding affinity based on the AlphaFold predicted structure.

## 1. Introduction

One conventional computational approach for drug design is virtual screening (VS).[1] However, this technology is limited by the small chemical space of existing compounds. It is reported that the number of drug-like molecules may be between $10^{23}$ and $10^{60}$,[2] while the one of registered compounds is only around $10^9$.[3] Recent years have witnessed rapid development of deep learning (DL)-based de novo molecular design. By DL-based generative models, molecules with specific properties are created.[4] One significant advantage of it is that large chemical space containing unregistered molecules can be explored or exploited.

In earlier stage, the DL generative algorithm for de novo drug design only depend on the knowledge of the structures of a certain amount of experimentally active compounds, namely ligand-based method.[5] Nonetheless, the molecular activities are determined by the interaction between the ligand and target, mainly one protein. In order to take advantage of the interaction information, in recent two years, both the structures of targes and ligands were considered to develop generative algorithms, called structure-based de novo methods.[6, 7] Taken on kind of reinforcement learning (RL) method for example, a generative model (agent) learns a policy (series of actions to create new structures) to generate molecules for maximizing a score, which is usually calculated by a predefined score function. Here, the score function includes the interaction between the ligand and target, such as molecular docking.[8] In addition, other DL algorithms have been used to create novel molecules such as recurrent neural network (RNN), autoencoder, generative adversarial network.

Despite some effort made in recent years, two limitations still hinder the practical application for de novo molecular design. One the one hand, most structure-based drug design method need experimental structures of the ligand-target complex as a starting point for molecular generation. Nowadays. ~18% of the total residues in human protein sequences are covered by experimentally determined structures. The experimental structures of ligand-target complexes are much fewer compared with the one of proteins themselves. Therefore, present structure-based algorithms are impractical to a high proportion of the human targets. On the other hand, in each step of molecular generation, the affinities between the given protein and new molecules should be simulated. Nevertheless, most of these processes is computationally intensive due to expensive 3D descriptors. Additionally, the problem is exacerbated when sampling in a substantial number of steps is indispensable to find the specific chemical space.

Guo et al. used curriculum learning to reduce the number of training steps, but the burden in each step has not been alleviated.

Besides 3D geometries of target binding pockets, sequence information of both molecules and proteins are collected as well as bioactive properties in some databases, such as Davis, Kiba and BindingDB. Thanks to the increase of this type of data, promising performance has been achieved to simulate drug-target affinity (DTA) by DL architectures using only 1D sequence of both molecules and proteins as inputs. Moreover, by this strategy, the interaction modeling is much more computationally efficient. Recently, Born et al. developed a RL method to design active compounds in a case study on SARS-CoV-2. Its score function included one of the DTA predictive models mentioned above. It proposed an effective approach for DTA prediction, which accelerate the interaction prediction in each step. However, three shortcomings still exist in this work: 1) 1D protein sequence-based DTA model does not belong to conventional pharmacophore or structure-based activity evaluation. However, neither ligand-based nor docking simulation has done to collect further evidence that the molecules are likely to be bioactive; 2) the drug-target interaction predictor is trained with the BindingDB dataset, it is unknown whether the predictors trained from different data source affect the performance of the generated model; 3) Except target related to SARS-CoV-2, other kinds of targets has not been studied using this type of generation models.

Herein, we proposed a widely used and fast protein sequence-based RL model for de novo drug design in which one score component is the function of DTA. Firstly, for obtaining binding predictive function, convolutional neural networks were trained by 1D protein sequence and molecular SMILES from three data sources, Davis, Kiba and BindingDB, respectively. Then, we generated potential inhibitors by our strategies toward epidermal growth factor receptor (EGFR), Glycogen synthase kinase-3 beta (GSK3β), serotonin 1A receptor (5HT1A) and dopamine D3 receptor (D3R). Regarding the evaluation metrics from the benchmark, the molecular diversity depends on the selection of the predictive function from different data sources. Next, given the analysis of QSAR and molecular docking, the potential bioactivities of the generated compounds were found. Finally, de novo drug design for a kinase without any experimental structure, CDK20, was analyzed. The novel molecules designed by our strategy showed similar pharmacophore features with a few first-in-class hit molecules, as well as favorable docking score based on the AlphaFold simulated structure.

## 2. Result and discussion

### 2.1 Performance and rate of the DTA predictive model

In our predictive model, two CNN-blocks learn the knowledge of molecules and protein, respectively. Then, two latent representations are concatenated into a fully-connected network for the interaction modeling. In the training process, Hyperparameter optimization was performed by grid search. The training method in detail was shown in **Method**. Table 1 illustrates the best result of DTA predictor for 3 datasets as well as corresponding architecture. The model trained from Kiba dataset achieved the best performance based on the MSE (0.153) and $R^2$ (0.775), while the predictor trained from BindingDB performed worsen than other two ones. The result above is consistent with previous work. To our knowledge, BindingDB dataset include various types of targets, including kinases, G protein-coupled peceptors (GPCRs) and others. In contrast, Kiba and Davis only contain kinase targets. Hence, more complicated composition might result in more difficulties to learn the DTA for the model. However, considering that MSE from BindingDB is less than 0.8, it means the error is smaller than one magnitude for the bioactivity prediction. Consequently, all the three models are assumed to be able to predict the DTA efficiently.

As for the inference time, it takes less than 20 ms for the prediction of 5000 molecules by three models. The time differences arise from the structure variance among three networks. On the contrary, Glide, theconventional molecular docking software, costs around 45 min for the prediction of the same amount of molecules, which even exclude the time spend for the generation of molecular 3D structures. It indicates that the pace of molecular generation will benefit from our strategy significantly, due to the great decrease of the DTA prediction in each RL step.

**Table 1.** The hyperparameters, performance and inference time of DTA predictive model.

|  | Windows | Max_seq | Max_smi | MSE | MSE_std | $R^2$ | $R^2$_std | Time (ms) |
| --- | --- | --- | --- | --- | --- | --- | --- | --- |
| BindingDB | 64 | 1000 | 100 | 0.792 | 0.012 | 0.672 | 0.004 | 16.9 |
| Kiba | 32 | 1000 | 100 | **0.153** | 0.003 | **0.775** | 0.004 | **15.4** |
| Davis | 64 | 1200 | 85 | 0.196 | 0.006 | 0.694 | 0.007 | 17.9 |

## 2.2 The MOSES evaluation of the generated molecules

We first sampled 10,000 molecules using each DTA predictors as well as Glide. Then, the MOSES evaluation metrics is applied as a comprehensive benchmark. As shown in Table 2, we found that the generated molecules by three methods showed excellent and analogous novelty and validity, which can be explained by the same prior and agent we used. Nonetheless, there are considerable differences in the performance of internal diversity (internal diversity), similarity to a nearest neighbor (SNN) and filters. As for Intdiv and SNN, both represent diversity of the generated set. According to the result, the model using DTA from BindingDB performed best for the molecular diversity, while model from Davis achieved lowest diversity performance. As shown in Method, the number of compounds is 68, 2,052 and 83,679. Therefore, it is assumed that the molecular diversity of the dataset trained for DTA predictor leads to the different performance above. To strengthen our assumption, The chemical distributions of the molecules from different datasets are T-SNE plotted in Supplementary Figure 1&2. It can be seen that molecules from BindingDB occupied the largest area of the chemical space. Meanwhile, the space area of molecules from Kiba cover the one from Davis. Thus, the validity of our hypothesis was confirmed: high diversity of the dataset trained for DTA model results in great variety of generated molecules. Regarding filters, the reason for the difference is similar to the one of diversity. Filters represents the fraction of the samples without unwanted fragments. The chemical space of the molecules from BindingDB is well-explored. Hence, maximum amount of reasonable molecules were generated by the model trained from BindingDB. Additionally, the molecules sampled by three different predictors The kernel density estimations of sampled molecules for the four properties (MW, SA, QED and LogP) were separately presented in Supplementary Fig. 1, which indicated that our method can generate molecules with high drugability.

**Table 2**. The MOSES evaluation metrics of the molecules generated using three DTA predictor and Glide.

|  | Filters | IntDiv | Novelty | SNN | Valid |
|---|---|---|---|---|---|
| BindDB | **0.906** | **0.811** | 0.989 | **0.370** | 0.976 |
| Kiba | 0.851 | 0.736 | **0.996** | 0.440 | 0.987 |
| Davis | 0.801 | 0.662 | 0.956 | 0.501 | **0.989** |

## 2.3 The pharmacophore-based activity evaluation for sampled molecules

In order to study the potential bioactivity of the generated molecules, pharmacophore-based studies were performed based on each QSAR model. Table 3 showed that the model from Kiba achieved best performance for GSK3β and EGFR targets, while the molecules sampled from BindingDB has the highest percentage of active compounds for 5HT1A and D3R targets. On the one hand, GSK3β and EGFR belong to tyrosine kinases. In RL sampling process, molecules are generated with high DTA predictive score for these two targets. Therefore, the model from Kiba is likely to generate high proportion of active compounds, because of high accuracy of the DTA prediction for kinase inhibitors. On the other hand, 5HT1A and D3R targets belong to GPCRs. BindingDB inludes GPCR targets and corresponding active molecules, while Kiba and Davis only focus on kinases. Consequently, it is assumed the DTA model based on Kiba and Davis has not learned the knowledge of ligand-GPCR target affinity. In other words, it is assumed that the generalization abilities of the two models are less accurate enough to predict the GPCR-related data. To strengthen our assumption, the modeling performance of the DTA model from Kiba and Davis was evaluated by the data of ligand-GPCR target. As a result, the $r^2$ were only 0.312 and 0.261 for Kiba and Davis, respectively. It implied that the generalization abilities of the two models were less accurate enough to predict the GPCR-related affinity, compared with the one from BindingDB. Hence, DTA predictor from BindingDB performed best for generating active GPCRs inhibitors. In contrast, the choice of generating active kinase and GPCRs inhibitors is DTA predictor training from Kiba and BindingDB, respectively.

**Table 3.** The ratio of the active compounds sampled by different models

|  | GSK3β | EGFR | 5HT1A | D3R |
|---|---|---|---|---|
| Random | 2.20 | 3.6 | 1.7 | 2.3 |
| Kiba | **12.4** | **47.8** | 2.35 | 4.1 |
| Davis | 8.5 | 22.5 | 2.52 | 2.1 |
| BindingBD | 3 | 10.4 | **7.3** | **9.8** |

Despite the different performance among the predictors, we found that most molecular set generated by RL methods has higher ratio of active compounds compared with randomly selected ones from CHEMBL. It indicated that our generative model can design molecules for specific target. To further study the chemical distributions of the generated molecules and existed active compounds, T-SNE plots of generated molecules and existing inhibitors were performed and shown in Figure 1. Here, the molecules are sample by RL using the predictor with the best performance (Kiba for GSK3β and EGFR, and BindingDB for 5HT1A and D3R). Firstly, in Figure 1, we found that partial area of the existing inhibitors for each target overlapped with sampled molecules by our method. It suggested that a part of generated molecules has high similarity with the inhibitors. Secondly, it is noticed that the overlapping areas are different among four cases. Specifically speaking, the area for EGFR is much larger than the ones for other three targets. It is consistent with the result of the ratio of active compound. As shown in Figure 1b, the inhibitors of EGFR occupied larger area of chemical space than other targets. Hence, the generated molecules are more likely to have high similarity with EGFR inhibitors than others. Thirdly, except for active compound evaluate by QSAR method, larger proportion of the sampled molecules have not overlap with the existing inhibitors. In other words, the potential activities are unknow for the molecules in this area when only use pharmacophore-based method. We will study their activity by molecular docking method in the next section.

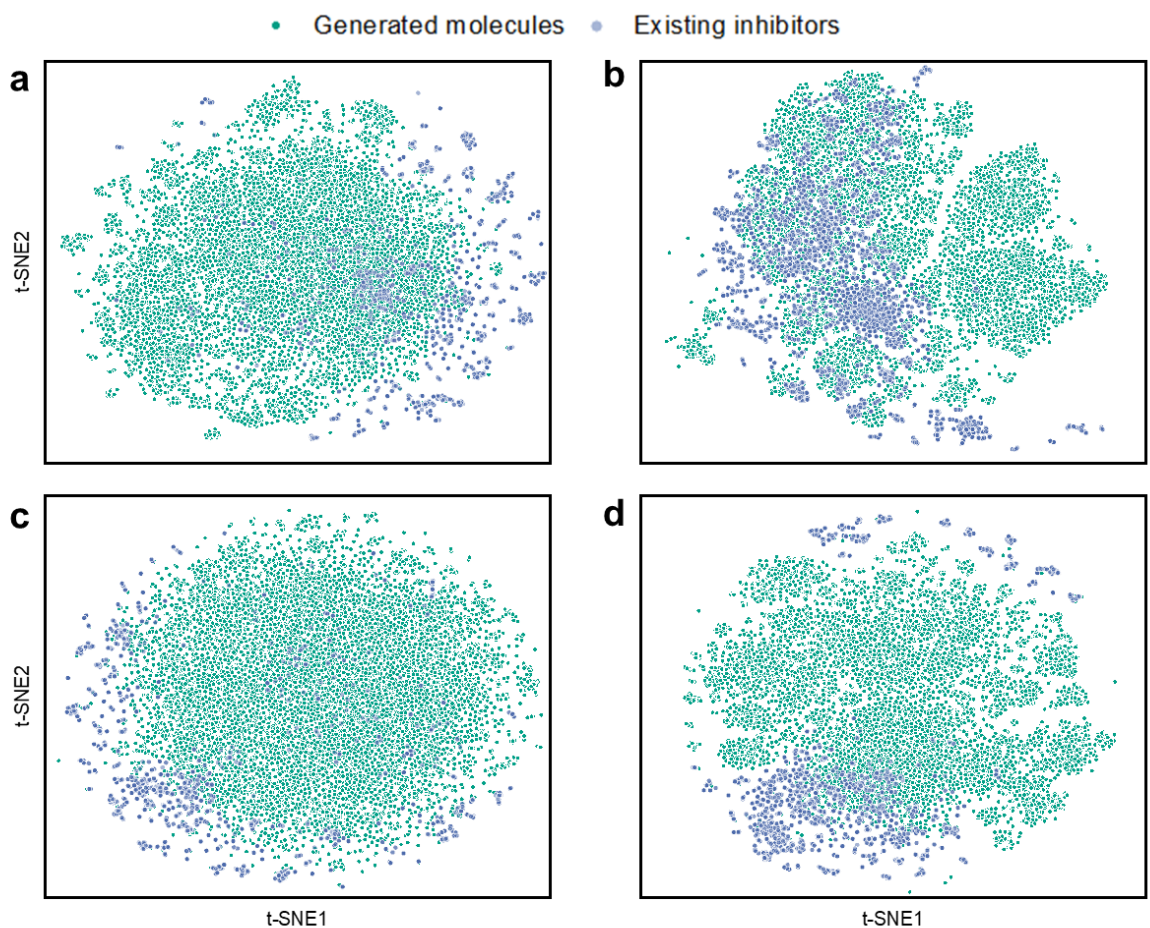

**Figure 1.** Two-dimensional t-SNE analysis of the the molecules generatedby by RL method using different predictor and existing inhibitors for GSK3β (a) and EGFR (b), 5HT1A (c) and D3R (d), respectively

## 2.4 Generating molecules without experimental protein structure
CDK20 has strong disease association, limited experimental structure information and with no publicly small molecule inhibitor. Three strategies were used to generate molecules for this target: DTA component, tanimoto similarity and the two above together. In Figure 2, The kernel density estimation distribution of the docking score measured by Glide showed that RL by DTA component perform best. It indicated that Protein Sequence-Based RL model can generate potential active compounds without 3D geometries of target binding pockets. The structures of the generated compounds with high docking score were shown in Figure 3.

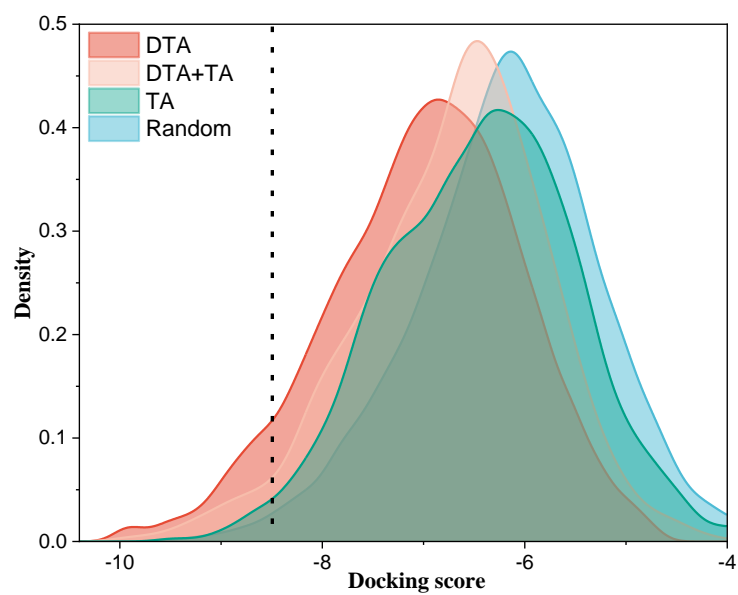

**Figure 3.** KDE (Kernel Density Estimation) distributions of the pharmacophore and docking scores of 10,000 molecules generated by DTA, tanimoto similarity and DTA+ tanimoto similarity

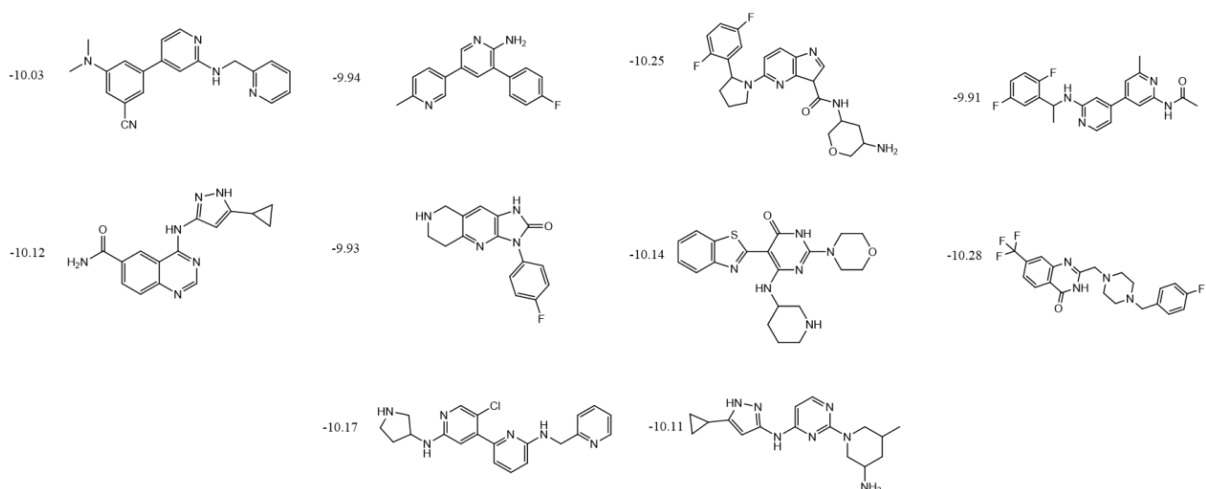

**Figure 4.** Potential active compounds with high docking score generated by DTA method for CDK20 target

## 3. Conclusion

We demonstrate a protein sequence-based RL model for widely used and fast de novo drug design. With 1D protein and molecule data as predictor input, potential active molecules are generated by a fast DTA model as one of the score components in RL process. Moreover, the bioactivities were confirmed by ligand-based QSAR and molecular docking methods. As for the ligand-based analysis, it is found that the ratio of active compounds depends on the selection of DTA predictor trained by different data source. By molecular docking study, it showed that our sampled molecules achieve high docking score with small similarity to the existing active inhibitors, which indicated that our generative model could explore the unknow chemical space. In addition, our strategy could generate more high docking-score molecules in less computational expense, compared to RL using external docking program. Finally, de novo design for a kinase without any experimental structure was studied by our protein sequence-based model. The sampled molecule showed potential bioactivity by comparing the docking score with the limited amount of experimentally active compounds based on AlphaFold predicted structure. Accordingly, our protein sequence-based RL model provided a new strategy to overcome the data scarcity of both experimental inhibitors and protein (or complex) structures, as well as expensive computational cost. From these results, de novo molecular design is anticipated to become a more practical method to facilitate drug discovery.